\documentclass[prl,reprint,superscriptaddress,preprintnumbers]{revtex4-2}

\newif\ifpublic\publictrue

\usepackage{fancyhdr}
\ifpublic\else
\fi
\newif\ifworking\workingtrue

\usepackage{mathtools,mathrsfs,amsbsy,amssymb,latexsym,amsfonts,amscd,
	amsmath, multirow}
\usepackage{bm}

\usepackage{graphicx}
\usepackage[usenames,dvipsnames]{color}
\usepackage[normalem]{ulem}
\usepackage{natbib}
\usepackage{xcolor}
\usepackage{comment}
\usepackage[normalem]{ulem}
\definecolor{linkcolor}{rgb}{0,0,0.6}
\usepackage[ pdftex,colorlinks=true,
pdfstartview=FitV,
linkcolor= linkcolor,
citecolor= linkcolor,
urlcolor= linkcolor,
hyperindex=true,
hyperfigures=false]
{hyperref}

\begin{document}
\title{Consistent $\mathcal{N}=4$, $D=4$ truncation of type IIB supergravity on $\textrm{S}^{1} \times \textrm{S}^{5}$}

\author{Adolfo Guarino}
\email{adolfo.guarino@uniovi.es}
\affiliation{Departamento de F\'isica, Universidad de Oviedo, Avda. Federico Garc\'ia Lorca 18, 33007 Oviedo, Spain}
\affiliation{Instituto Universitario de Ciencias y Tecnolog\'ias Espaciales de Asturias (ICTEA), Calle de la Independencia 13, 33004 Oviedo, Spain}

\author{Colin Sterckx}
\email{colin.sterckx@pd.infn.it}
\affiliation{INFN, Sezione di Padova,\\
Via Marzolo 8, 35131 Padova, Italy.}

\author{Mario Trigiante}
\email{mario.trigiante@polito.it}
\affiliation{Department of Applied Science and Technology, Politecnico di Torino, C.so Duca degli Abruzzi, 24, 10129, Turin, Italy.\\ INFN - Sez. Torino, Via P. Giuria, 1, 10125, Turin, Italy.}

\begin{abstract}

Fetching techniques from Generalised Geometry and Exceptional Field Theory, we develop a new method to identify consistent subsectors of four-dimensional gauged maximal supergravities that possess a (locally) geometric embedding in type IIB or 11D supergravity. We show that a subsector that is invariant under a structure group $\textrm{G}_\textrm{S} \subset \textrm{E}_{7(7)}$ can define a consistent truncation, even when $\textrm{G}_\textrm{S}$ is not part of the symmetry of the gauged maximal supergravity. As an illustration of the method, type IIB supergravity on $\textrm{S}^{1} \times \textrm{S}^{5}$ is shown to admit a consistent truncation to pure $\mathcal{N}=4$, $D=4$ gauged supergravity. Explicit uplift formulae are presented which provide a type IIB alternative to the M-theory embedding constructed by Cvetic, Lu and Pope $25$ years ago.

\end{abstract}

\pacs{04.65.+e, 04.50.+h,11.25Mj}

\maketitle

\section{Introduction}

Consistent truncations of 11D/type II supergravity down to \emph{pure} $\mathcal{N}$-extended $D=4$ gauged supergravity with $2 \le \mathcal{N}\le 4$ provide simple playgrounds for gauge-gravity applications and are essential for the holographic study of universal conformal field theory (CFT) observables \cite{Maldacena:1997re}. When containing AdS$_{4}$ vacua, such minimalist truncations solely retain the supergravity fields dual to the stress-tensor superconformal multiplet of a strongly-coupled three-dimensional CFT (in the large $N$ limit). Examples of consistent truncations with $2 \le \mathcal{N}\le 4$ have been constructed following two approaches. The first one is a top-down approach in which an educated guess is made about what the higher-dimensional fields might be in terms of the four-dimensional ones. A pioneering example was the truncation of 11D supergravity on $\textrm{S}^{7}$ down to the pure $\mathcal{N}=4$ $\textrm{SO}(4)_{\textrm{R}}$-gauged supergravity of \cite{Das:1977pu} worked out in \cite{Cvetic:1999au}. Following this top-down approach, other examples of consistent truncations of 11D supergravity include the truncation of \cite{Gauntlett:2007ma} on Sasaki--Einstein $7$-manifolds, or the truncation of \cite{Larios:2019lxq} on the seven-dimensional geometries of \cite{Gabella:2012rc}, all of them down to pure $\mathcal{N}=2$ gauged supergravity. Another instance is the truncation of \cite{Cassani:2011fu} on tri-Sasakian $7$-manifolds down to pure $\mathcal{N}=3$ $\textrm{SO}(3)_{\textrm{R}}$-gauged supergravity.

The second approach is purely four-dimensional and makes use of a pre-existing consistent truncation of 11D/type II supergravity down to a \textit{maximal} ${\mathcal{N}=8}$ gauged supergravity (which only consists of the $\mathcal{N}=8$ supergravity multiplet). Starting from the maximal theory, one proceeds to identify consistent subsectors that describe pure $\mathcal{N}$-extended gauged supergravities with $2 \le \mathcal{N}\le 4$. An example is the consistent truncation of 11D supergravity on $\textrm{S}^{7}$ \cite{deWit:1986oxb} down to the $\mathcal{N}=8$ supergravity with an $\textrm{SO}(8)$ gauging \cite{deWit:1982ig}, which contains two inequivalent subsectors describing pure $\mathcal{N}=2$ gauged supergravity \cite{Larios:2019kbw}. Each subsector provides a different embedding of pure $\mathcal{N}=2$ gauged supergravity in 11D. While one of them is underpinned by a standard symmetry principle (it corresponds to a specific $\textrm{SU}(4)_{s} \subset \textrm{SO}(8)$ invariant subsector of the maximal theory) and recovers the ansatz of \cite{Gauntlett:2007ma}, the other is not and provides a new truncation ansatz whose consistency must be checked at the level of the field equations. There is also the consistent truncation of massive IIA supergravity on $\textrm{S}^{6}$ \cite{Guarino:2015vca} down to the $\mathcal{N}=8$ supergravity with an $\textrm{ISO}(7)$ gauging \cite{Guarino:2015qaa}, which contains two subsectors describing pure $\mathcal{N}=2$ and $\mathcal{N}=3$ gauged supergravity \cite{Varela:2019vyd}. This time none of the two is based on a standard symmetry principle (they do not describe $\textrm{G}_{0}\subset \textrm{ISO}(7)$ invariant subsectors of the maximal theory), so their consistency was established again at the level of the field equations.

In all the cases not based on a standard symmetry principle, the first step to work out a consistent truncation to pure $\mathcal{N}$-extended gauged supergravity is the identification of an AdS$_{4}$ vacuum in the gauged maximal supergravity that preserves the same number $\mathcal{N}$ of supersymmetries. As a second step, the gauged maximal supergravity is judiciously expanded about such an $\mathcal{N}$-supersymmetric AdS$_{4}$ vacuum so that the field content of the pure $\mathcal{N}$-extended gauged supergravity is identified within the maximal theory and retained, thus providing a description of the dynamics of the maximal theory about that vacuum. Due to the lack of a standard symmetry principle, one must verify that this field content describes a consistent subsector of the maximal theory at the level of the field equations. In a last step, the field content of the pure $\mathcal{N}$-extended supergravity is oxidised to 11D/type II supergravity using general uplift formulae available for the gauged maximal supergravity. With the advent of the Exceptional Field Theory \cite{Hohm:2013pua,Hohm:2013uia} and its generalised Scherk--Schwarz reductions \cite{Hohm:2014qga}, there is by now a systematic method to derive such uplift formulae for those gauged maximal supergravities that possess a higher-dimensional origin \cite{Inverso:2017lrz}. 

In this letter we will argue that all the cases in the second approach lacking a standard symmetry principle can still be understood on the basis of a symmetry principle, and their consistency can be established without having to invoke equations of motion. Our main result can be stated as:
\\[-3mm]

\noindent\textit{``Any $D$-dimensional $\mathcal{N}$-extended supergravity, which can be realised via a compactification that is locally described by ten- or eleven-dimensional maximal supergravity and which exhibits an AdS$_D$ or Mkw$_D$ solution preserving $\mathcal{N}'\leq \mathcal{N}$ supersymmetries, contains a consistent subsector that describes pure $D$-dimensional $\mathcal{N}'$-extended supergravity. Such a consistent subsector is invariant under the maximal structure group $\textrm{G}_S$ leaving the $\mathcal{N}'$ preserved supercharges invariant."}
\\[-3mm]

\noindent Note that the group $\textrm{G}_S$ does \textit{not} need to be a subgroup of the symmetry group of the $\mathcal{N}$-extended supergravity. The proof  of consistency solely relies on generalised geometry and does not require the verification of the field equations. Our result can be viewed as the lower-dimensional analogue of the proof of \cite{Gauntlett:2007ma} provided in \cite{Cassani:2019vcl}.

As an illustration of the above, we will consider the maximal $\mathcal{N}=8$ supergravity of \cite{Inverso:2016eet} which arises from an S-fold compactification of type IIB supergravity on $\textrm{S}^{1} \times \textrm{S}^{5}$ and contains an AdS$_{4}$ vacuum that preserves $\mathcal{N}'=4$ supersymmetry. We will identify a consistent subsector of the maximal theory that describes the pure $\mathcal{N}=4$ $\textrm{SO}(4)_{\textrm{R}}$-gauged supergravity of \cite{Das:1977pu}. Upon oxidation of this subsector back to ten dimensions, a novel type IIB embedding of pure $\mathcal{N}=4$ $\textrm{SO}(4)_{\textrm{R}}$-gauged supergravity can be obtained which provides an alternative to the celebrated M-theory embedding of \cite{Cvetic:1999au}.

\section{$\textrm{SO}(4)_{\textrm{R}}$-gauged  $\mathcal{N}=4$, $D=4$ supergravity}
\label{sec:N=4&D=4_sugra}

The bosonic sector of the $\mathcal{N}=4$, $D=4$ supergravity  multiplet consists of the spin-$2$ metric field $g_{\mu\nu}$, six spin-$1$ electric fields $A_{\mu}{}^{\Lambda}$ and their magnetic duals $\tilde{A}_{\mu \Lambda}$, with $\Lambda=1,\ldots,6$, and a spin-$0$ complex scalar $\tau$ that parameterises the coset space $\textrm{SL}(2)/\textrm{SO}(2)$. In addition, there are four spin-$3/2$ gravitini fields $\psi^{i}_{\mu}$ and four spin-$1/2$ fermions $\lambda^{i}$, with $i=1,\ldots,4$, transforming under the R-symmetry group $\textrm{SU}(4)_{\textrm{R}}$ of the theory. In its ungauged version, the theory possesses a global duality group $\mathcal{G}_{\mathcal{N}=4}=\textrm{SL}(2) \times \textrm{SO}(6)$ although only a subgroup $\mathcal{G}_{\mathcal{L}}$ is realised in the Lagrangian. In its gauged version, the vectors span a non-abelian gauge symmetry
\begin{equation}
\label{G_N4}
\textrm{G}_{\mathcal{N}=4}=\textrm{SO}(4)_{\textrm{R}} \subset \textrm{SU}(4)_{\textrm{R}} \ ,   
\end{equation}
under which the fermions and vectors transform but $\tau$ does not.

In the standard symplectic frame formulation of \cite{Schon:2006kz}, it was shown in \cite{Louis:2014gxa} that an electric gauging of (\ref{G_N4}) cannot accommodate an $\mathcal{N}=4$ AdS$_{4}$ vacuum. Instead, one needs both electric and magnetic vectors to enter the gauging simultaneously. The presence of magnetic vectors in the gauge connection requires the inclusion of auxiliary two-form tensor fields when it comes to write down a gauge invariant Lagrangian \cite{deWit:2005ub}. In order to avoid these additional complications, it will prove convenient to abandon the standard frame of \cite{Schon:2006kz} and perform a symplectic rotation so that the gauging becomes electric in the new symplectic frame. The symplectic rotation bringing the standard frame of \cite{Schon:2006kz} to the new one that we will use in this letter has been carefully discussed in \cite{DallAgata:2023ahj}. We refer the reader there for more details.

The bosonic action in the new symplectic frame takes the form
\begin{equation}
\label{action_N=4}
\begin{array}{rcl}
S_{\textrm{bos}} &=& \displaystyle \int (R- V) \star 1 - \tfrac{1}{2} \, d\xi \wedge \star d\xi - \tfrac{1}{2} \, e^{2\xi} \, d\chi \wedge \star d\chi  \\[2mm]
&  & - \, \dfrac{1}{1 + \chi^2 e^{2\xi}} \left( e^{\xi} F^{a}_{(2)} \wedge \star F^{a}_{(2)} + \chi \, e^{2\xi} \, F^{a}_{(2)} \wedge F^{a}_{(2)} \right) \\[4mm]
&  & - \,e^{-\xi} F^{\hat{a}}_{(2)} \wedge \star F^{\hat{a}}_{(2)} + \chi \, F^{\hat{a}}_{(2)} \wedge F^{\hat{a}}_{(2)}   \ ,
\end{array}
\end{equation}
in terms of the complex scalar $\tau \equiv -\chi + i e^{-\xi}$ and the non-abelian field strengths
\begin{equation}
\label{F_non_abelian}
\begin{array}{rcll}
F_{(2)}^{a}&=&dA_{(1)}^{a}+\frac{1}{2} \, g \, \epsilon_{abc} \, A_{(1)}^{b} \wedge A_{(1)}^{c} & , \\[2mm]
F_{(2)}^{\hat{a}} &=& dA_{(1)}^{\hat{a}}+\frac{1}{2} \, g \, \epsilon_{\hat{a}\hat{b}\hat{c}} \, A_{(1)}^{\hat{b}} \wedge A_{(1)}^{\hat{c}} & .
\end{array}
\end{equation}
The gauging also induces a scalar potential for $\tau$ of the form
\begin{equation}
\label{V_N4}
V = - g^{2} (4 + 2 \cosh \xi + e^\xi \chi^2) \ ,
\end{equation}
which possesses an AdS$_{4}$ vacuum at $\tau=i$ that preserves the $\textrm{SO}(4)_{\textrm{R}}$ gauge symmetry and $\mathcal{N}=4$ supersymmetries.

The bosonic action (\ref{action_N=4}) matches the one in \cite{Cvetic:1999au}\footnote{Fields are related by $\chi_{\cite{Cvetic:1999au}} = -\chi_{\text{here}}$, $\phi_{\cite{Cvetic:1999au}}= \xi_{\text{here}}$, $A_{\cite{Cvetic:1999au}} = \sqrt{2} A_{\text{here}}$ (for all vectors) and $g_{\cite{Cvetic:1999au}} = g_{\text{here}}/\sqrt{2}$.} (see \cite{Das:1977pu,Freedman:1978ra} for original references). This action was shown in \cite{Cvetic:1999au} to arise from a consistent truncation of 11D supergravity on $\textrm{S}^{7}$ or, alternatively, from a singular limit of type IIA supergravity on $\textrm{S}^{3} \times \textrm{S}^{3}$. In this letter we will argue that it also arises from a consistent truncation of type IIB supergravity on $\textrm{S}^{1} \times \textrm{S}^{5}$, hence providing a new and independent ten-dimensional embedding of the $\textrm{SO}(4)_{\textrm{R}}$-gauged $\mathcal{N}=4$, $D=4$ supergravity.

\section{$\mathcal{N}=4$ S-fold and consistent truncations}

Type IIB supergravity admits a class of (globally) non-geometric solutions, known as S-folds (see \cite{Guarino:2022tlw} for a review), which are a warped product of the form $\textrm{AdS}_{4} \times \textrm{S}^{1} \times \textrm{S}^{5}$. The solutions incorporate an S-duality twist along the $\textrm{S}^{1}$ which induces a non-trivial $\textrm{SL}(2,\mathbb{Z})$ monodromy of hyperbolic type on the type IIB fields when looping around the $\textrm{S}^{1}$. A pioneering example of these non-geometric (still locally geometric) backgrounds is the $\mathcal{N}=4$ supersymmetric S-fold of \cite{Inverso:2016eet} which features an $\textrm{SO}(4)_{\textrm{diag}} \sim \textrm{SO}(3)_{1} \times \textrm{SO}(3)_{2}$ symmetry that is realised geometrically on the internal $\textrm{S}^{5}$ when viewed as a product of two $2$-spheres $\textrm{S}^{2}_{i=1,2}$ fibered over an interval $I$. Building upon \cite{Duff:1985jd,Pope:1987ad}, it was conjectured in \cite{Gauntlett:2007ma} and more recently proved in \cite{Cassani:2019vcl} that:
\\[-3mm]

\noindent\textit{``Any supergravity solution with a $D$-dimensional AdS (or Minkowski) factor preserving $\mathcal{N}$ supersymmetries, defines a consistent truncation to the corresponding pure supergravity theory."}
\\[-3mm]

\noindent The above statement then implies the existence of a consistent truncation of type IIB supergravity on $\textrm{S}^{1} \times \textrm{S}^{5}$ down to pure $\mathcal{N}=4$, $D=4$ supergravity, where the four-dimensional avatar of the $\mathcal{N}=4$ supersymmetric S-fold of \cite{Inverso:2016eet} is the $\mathcal{N}=4$ AdS$_{4}$ extremum of the scalar potential (\ref{V_N4}) at $\tau=i$. We will show that this is indeed the case.

From now on we will frame our discussion within the well-established consistent truncation of type IIB supergravity on $\textrm{S}^{1}\times\textrm{S}^{5}$ down to the gauged maximal supergravity of \cite{Inverso:2016eet} with gauge group
\begin{equation}
\label{Gauge_group}
\textrm{G}_{\mathcal{N}=8}=\left[\textrm{SO}(6)_{\textrm{diag}} \times \textrm{SO}(1,1) \right] \ltimes \mathbb{R}^{12} \ .
\end{equation}
This maximal supergravity possesses an AdS$_{4}$ vacuum that preserves $\mathcal{N}=4$ supersymmetries and $\textrm{SO}(4)_{\textrm{diag}} \subset \textrm{SO}(6)_{\textrm{diag}} \subset \textrm{G}_{\mathcal{N}=8}$ \cite{Gallerati:2014xra}, and which was uplifted to the $\mathcal{N}=4$ S-fold background of type IIB supergravity in \cite{Inverso:2016eet}. We will use this $\mathcal{N}=4$ AdS$_4$ vacuum of the gauged maximal supergravity of \cite{Inverso:2016eet} as an anchor to work out the consistent truncation of type IIB supergravity down to the pure $\mathcal{N}=4$ and $\textrm{SO}(4)_{\textrm{R}}$-gauged supergravity of Sec.~\ref{sec:N=4&D=4_sugra}.

\subsection{An unorthodox $\mathcal{N}=4$ subsector}
\label{sec:unorthodox_sector}

First of all, one may wonder whether there is a group-theoretically consistent subsector of the gauged maximal supergravity of \cite{Inverso:2016eet} that describes pure $\mathcal{N}=4$ supergravity and contains the $\mathcal{N}=4$ AdS$_{4}$ vacuum. Using representation theory, it was shown in \cite{Guarino:2024zgq} that such a subsector does not exist. The reason is the following. The $\mathcal{N}=4$ AdS$_{4}$ vacuum defines an $\textrm{SU}(4)_{\textrm{S}}$ structure. Namely, it selects an $\textrm{SU}(4)_{\textrm{S}}$ subgroup inside the $\textrm{SU}(8)_{\textrm{R}}$ R-symmetry group that leaves invariant the four supercharges preserved at the $\mathcal{N}=4$ vacuum. It also selects a commuting $\textrm{U}(4)_{\textrm{R}}$ inside $\textrm{SU}(8)_{\textrm{R}}$ that rotates the four supercharges. Therefore, defining an $\textrm{SU}(4)_{\textrm{S}}$ structure amounts to specifying an embedding 
\begin{equation}
\label{SU(4)_S_embedding}
\textrm{SU}(4)_{\textrm{S}} \times \textrm{U}(4)_{\textrm{R}} \subset \textrm{SU}(8)_{\textrm{R}} \ ,
\end{equation}
with $\textrm{U}(4)_{\textrm{R}} = \textrm{SU}(4)_{\textrm{R}} \times \textrm{U}(1)_{\textrm{R}}$. The field content of pure $\mathcal{N}=4$ supergravity is then recovered as the $\textrm{SU}(4)_{\textrm{S}}$-invariant subsector of maximal supergravity. However, the $\mathcal{N}=4$ AdS$_{4}$ vacuum of the gauged maximal supergravity of \cite{Inverso:2016eet} preserves an $\textrm{SO}(4)_{\textrm{diag}}\subset \textrm{SU}(4)_{\textrm{diag}} \sim \textrm{SO}(6)_{\textrm{diag}} $ subgroup of (\ref{Gauge_group}) that is diagonally embedded as
\begin{equation}
\label{SU(4)_embedding_N=4_vacuum}
\textrm{SU}(4)_{\textrm{diag}} \subset \textrm{SU}(4)_{\textrm{S}} \times \textrm{SU}(4)_{\textrm{R}} \subset \textrm{SU}(8)_{\textrm{R}} \ .
\end{equation}
Since the $\bf{8}$ gravitini of maximal supergravity decompose as $\bf{8} = \left(\bf{\frac{1}{2}},\bf{\frac{1}{2}}\right) \oplus \left(\bf{\frac{1}{2}},\bf{\frac{1}{2}}\right)$ under $\textrm{SO}(4)_{\textrm{diag}}$, there is no subgroup of $\textrm{SO}(4)_{\textrm{diag}}$ that leaves invariant the four gravitini corresponding to the preserved supercharges at the $\mathcal{N}=4$ vacuum and projects out the other four.

We must therefore resort to the procedure sketched in the introduction to work out a consistent truncation of the maximal supergravity of \cite{Inverso:2016eet} about its $\mathcal{N}=4$ AdS$_{4}$ vacuum. The original gauging data of \cite{Inverso:2016eet} is encoded in a so-called embedding tensor $X$ which lives in the $\bf{912}$ irreducible representation (irrep) of $\textrm{E}_{7(7)}$ and codifies all the couplings amongst the various fields in the gauged maximal supergravity. Let us denote by
\begin{equation}
\mathcal{V}_{\textrm{vac}}(\Phi_{0}) \in \textrm{E}_{7(7)}/\textrm{SU}(8)   
\end{equation}
the scalar coset representative when evaluated at the $\mathcal{N}=4$ AdS$_{4}$ vacuum configuration $\left\langle \Phi_{0} \right\rangle$. It turns out that $\mathcal{V}_{\textrm{vac}}(\Phi_{0}) \neq \mathbb{I}$, so the set of consistent fluctuations about the AdS$_{4}$ vacuum that are compatible with its $\mathcal{N}=4$ supersymmetries must be carefully identified. Fortunately, the $\textrm{SU}(4)_{\textrm{S}}$ structure defined at the $\mathcal{N}=4$ vacuum does it for us. It automatically selects a unique $\textrm{SL}(2)$, \textit{i.e.} it defines an axio-dilaton structure \cite{Malek:2017njj}, through the embedding
\begin{equation}
\label{axiondilaton_structure}
\textrm{SU}(4)_{\textrm{S}} \times \textrm{SU}(4)_{\textrm{R}} \times \textrm{SL}(2) \subset \textrm{E}_{7(7)} \ ,
\end{equation}
so that the scalars $(\chi,\xi)$ of the pure $\mathcal{N}=4$ supergravity are the ones being invariant under $\textrm{SU}(4)_{\textrm{S}}$ (and also under $\textrm{SU}(4)_{\textrm{R}}$ by virtue of representation theory). These scalars parameterise the coset representative
\begin{equation}
\label{coset_SL(2)}
\mathcal{V}(\chi,\xi) \in\textrm{SL}(2)/\textrm{SO}(2) \ ,
\end{equation}
which, in our construction, encodes the scalar fluctuations about the $\mathcal{N}=4$ vacuum. As a result, the theory of \cite{Inverso:2016eet} specified by the embedding tensor $X$ must be expanded about its $\mathcal{N}=4$ vacuum using the coset representative
\begin{equation}
\label{coset_fluc}
\mathcal{V}_{\textrm{fluc}}(\chi,\xi) =\mathcal{V}_{\textrm{vac}}(\Phi_{0}) \cdot \mathcal{V}(\chi,\xi)\ .
\end{equation}
Introducing indices $\hat{\imath} = 1 ,\ldots , 4$ for $\textrm{SU}(4)_{\textrm{S}}$ and $i = 1 ,\ldots , 4$ for $\textrm{SU}(4)_{\textrm{R}}$ in the fundamental representations, the electric and magnetic vectors of the pure $\mathcal{N}=4$ supergravity are identified with the vectors $A_{\mu}{}^{[ij]}$ and $A_{\mu}{}_{[ij]}$ in the gauged maximal supergravity. This can be seen by branching the $\bf{56}$ of $\textrm{E}_{7(7)}$ (where vectors live) under (\ref{axiondilaton_structure}).

An alternative (but equivalent) viewpoint can be adopted. Since the expansion of the gauged maximal supergravity takes place at $\mathcal{V}_{\textrm{vac}}(\Phi_{0}) \neq \mathbb{I}$, one can consider the local form of the embedding tensor (also known as $T$-tensor) at the $\mathcal{N}=4$ vacuum. This is given by
\begin{equation}
\label{X_new}
\widetilde{X} = \mathcal{V}^{-1}_{\textrm{vac}}(\Phi_{0})  \star  X \ ,
\end{equation}
where $\star$ denotes the left action of the $\textrm{E}_{7(7)}$ element $\mathcal{V}_{\textrm{vac}}(\Phi_{0})$ on the original embedding tensor $X \in \bf{912}$ of \cite{Inverso:2016eet}. The $\widetilde{X}$ in (\ref{X_new}) then codifies all the couplings in the maximal gauged supergravity of \cite{Inverso:2016eet} once it has been Higgsed and expanded about its $\mathcal{N}=4$ vacuum. 

Equipped with the original embedding tensor $X$ of \cite{Inverso:2016eet} and the coset representative $\mathcal{V}_{\textrm{fluc}}(\chi,\xi)$ in (\ref{coset_fluc}), one can determine the theory describing the dynamics of the fluctuations $(\chi,\xi)$ about the $\mathcal{N}=4$ AdS$_{4}$ vacuum.  Such dynamics is encoded in the two \emph{fermion shift tensors}, or, simply, \emph{fermion shifts},  $\mathcal{A}_{1}$ and $\mathcal{A}_{2}$ (and their complex conjugates)  \cite{deWit:2007mt}, that transform under $\textrm{SU}(8)_{\textrm{R}}$ in the $\bf{36}$ and $\bf{420}$ irreps (and the conjugate representations), respectively. On a given vacuum, the two tensors $\mathcal{A}_1,\, \mathcal{A}_2$ encode the gravitini and the dilatini masses, respectively. A standard computation of $\mathcal{A}_1$ and $\mathcal{A}_2$ in the maximal theory, about the $\mathcal{N}=4$ vacuum, yields the following non-vanishing components:
\begin{equation}
\label{A1_tensor}
(\mathcal{A}_{1})_{\hat{\imath} \hat{\jmath}} = 2 f_{+} \, \delta_{\hat{\imath} \hat{\jmath}}
\hspace{4mm} , \hspace{4mm}
(\mathcal{A}_{1})_{ij} = \bar{f}_{+}  \,  \delta_{ij} \ ,
\end{equation}
and
\begin{equation}
\label{A2_tensor}
\begin{array}{c}
(\mathcal{A}_{2})_{\hat{\imath}}{}^{\hat{\jmath} kl} = -2 f_{+} \, \delta_{\hat{\imath}\hat{\jmath}}^{kl}+ f_{-} \, \epsilon^{\hat{\imath}\hat{\jmath}kl}
\hspace{3mm} , \hspace{3mm}
(\mathcal{A}_{2})_{\hat{\imath}}{}^{\hat{\jmath}\hat{k} l} = i\, \sqrt{2} \, \epsilon^{\hat{\imath} \hat{\jmath} \hat{k} l} \ , \\[2mm]
(\mathcal{A}_{2})_{\hat{\imath}}{}^{\hat{\jmath}\hat{k}\hat{l}} = 2\bar{f}_{-}\, \epsilon^{\hat{\imath}\hat{\jmath}\hat{k}\hat{l}}
\hspace{3mm} , \hspace{3mm}
(\mathcal{A}_{2})_{i}{}^{jkl}= {f}_{-} \, \epsilon^{ijkl} \ ,
\end{array}
\end{equation}
where we have introduced the scalar-dependent functions
\begin{equation}
f_{\pm}(\chi,\xi) \equiv \tfrac{1}{2} \, e^{\frac{\xi }{2}} \left(i \chi +e^{-\xi} \pm 1\right) \ .
\end{equation}
Note that the fermion shifts in (\ref{A1_tensor}) and (\ref{A2_tensor}) are manifestly invariant under the diagonal $\textrm{SO}(4)_{\textrm{diag}} \subset \textrm{SU}(4)_\textrm{diag}$ specified in (\ref{SU(4)_embedding_N=4_vacuum}) which identifies $\hat{\imath}$ with $i$. As a check of consistency, the scalar potential computed from the fermions shifts in (\ref{A1_tensor}) and (\ref{A2_tensor}) reproduces the one in (\ref{V_N4}). When particularised to the $\mathcal{N}=4$ AdS$_{4}$ vacuum, \textit{i.e.} at $\chi=\xi=0$, one finds that $(f_{+},f_{-})=(1,0)$, $L=g^{-1}$ becomes the AdS$_{4}$ radius, and the $\mathcal{N}=4$ preserved supersymmetries are associated with the four gravitini $\psi^{i}_{\mu}$ (with unit normalised mass in (\ref{A1_tensor})) transforming under the R-symmetry group $\textrm{SU}(4)_{\textrm{R}}$ of $\mathcal{N}=4$ supergravity.

\subsection{A proof of consistency}

The consistency of the unorthodox $\mathcal{N}=4$ subsector describing pure $\mathcal{N}=4$ supergravity can be proved along the lines of \cite{Malek:2017njj,Cassani:2019vcl} on the basis of the existence of a generalised $\textrm{SU}(4)_{\textrm{S}}$ structure whose generalised intrinsic torsion $T^{\textrm{int}}_{\mathcal{N}=4}$ is a constant $\textrm{SU}(4)_{\textrm{S}}$ singlet (see \cite{Sterckx:2024vju} for a recent review). The generalised identity structure underlying the consistent truncation of type IIB supergravity on $\textrm{S}^{1} \times \textrm{S}^{5}$ down to the gauged maximal supergravity of \cite{Inverso:2016eet} (which retains all the fields in the $\mathcal{N}=8$, $D=4$ supergravity multiplet) uniquely determines the generalised $\textrm{SU}(4)_{\textrm{S}}$ structure responsible for the consistency of the unorthodox $\mathcal{N}=4$ subsector through the group-theoretic embedding in (\ref{axiondilaton_structure}).

Denoting by $E$ the bundle associated with the $\bf{56}$ of $\textrm{E}_{7(7)}$,  the intrinsic torsion of an $\textrm{SU}(4)_{\textrm{S}}$ structure  is defined as the quotient
\begin{equation}
\label{W_int}
W^{\textrm{int}} = W/\textrm{Im}\left(\tau_{\textrm{SU}(4)_{\textrm{S}}}\right) \ ,
\end{equation}
where $W = \mathbf{912} \subset E^{*} \otimes \mathfrak{e}_{7(7)}$ is the space of generalised intrinsic torsions for identity structures, which encode the embedding tensor of maximal supergravity, and $\tau_{\textrm{SU}(4)_{\textrm{S}}}$ is the projection of the space of  $\textrm{SU}(4)_{\textrm{S}}$-compatible connections $K = E^{*} \otimes \mathfrak{su}(4)_{\textrm{S}}$ into $W$.  A simple representation theory exercise based on the chain of embeddings
\begin{equation}
\textrm{E}_{7(7)} \supset  \textrm{SU}(8)_{\textrm{R}} \supset  \textrm{SU}(4)_{\textrm{S}}  \times  \textrm{SU}(4)_{\textrm{R}}  \times \textrm{U}(1)_{\textrm{R}}  \ ,
\end{equation}
shows that the space of torsions $W = \mathbf{912}$ decomposes as
\begin{equation}
\label{W_irreps}
\begin{array}{rcl}
\mathbf{36} &\rightarrow& (\mathbf{1},\,\mathbf{10})_{-2} \oplus (\mathbf{4},\,\mathbf{4})_0 \oplus (\mathbf{10},\,\mathbf{1})_{2} \ , \\[2mm]
\mathbf{420} &\rightarrow& (\mathbf{1},\,\mathbf{\bar{10}})_{-2}\oplus(\mathbf{4},\,\mathbf{4})_{0} \oplus(\mathbf{\bar{10}},\,\mathbf{1})_{2} \,  \oplus   \\[1mm]
& &  (\mathbf{15},\,\mathbf{6})_{2} \oplus (\mathbf{6},\,\mathbf{15})_{-2} \oplus(\mathbf{20},\,\mathbf{4})_{0} 
\oplus(\mathbf{4},\,\mathbf{20})_{0} \,  \oplus \\[1mm]
& & (\mathbf{\bar{4}},\,\mathbf{\bar{4}})_{\pm 4} \oplus(\mathbf{6},\,\mathbf{1})_{2} \oplus(\mathbf{1},\,\mathbf{6})_{-2}  \ ,
\end{array}
\end{equation}
together with the complex conjugated (c.c.) irreps. The analysis of $K$ gives
\begin{equation}
\label{K_irreps}
\begin{array}{rcl}
K & \subset &  \big[(\mathbf{1},\mathbf{6})_{-2}\oplus (\mathbf{4},\mathbf{4})_0 \oplus(\mathbf{6},\mathbf{1})_2  \oplus \textrm{c.c.} \big] \otimes  (\mathbf{15},\mathbf{1})_0  \\[2mm]
& = & (\mathbf{15},\,\mathbf{6})_{-2} \oplus (\mathbf{36} \oplus \mathbf{20} \oplus \mathbf{4},\,\mathbf{4})_0 \,  \oplus \\[2mm]
&    & (\mathbf{64}\oplus \mathbf{\bar{10}} \oplus \mathbf{10} \oplus \mathbf{6},\,\mathbf{1})_2 \oplus \textrm{c.c.}
\end{array}
\end{equation}

The intrinsic torsion $T^{\textrm{int}}_{\mathcal{N}=4}$ is identified with $\tilde{X}$ in (\ref{X_new}) mod-out by $K$. Using the branching rules in (\ref{W_irreps}), the components of $\widetilde{X}$ can be read off from the fermion shifts in (\ref{A1_tensor}) and (\ref{A2_tensor}) evaluated at the $\mathcal{N}=4$ vacuum, \textit{i.e.} at $\chi=\xi=0$. Then, the well-established type IIB embedding of the gauged maximal supergravity of \cite{Inverso:2016eet}, combined with the systematics of consistent truncations developed in \cite{Cassani:2019vcl}, ensure that the $\mathrm{SU}(4)_{\textrm{S}}$-singlets $(\chi,\xi)$ will describe a consistent type IIB truncation to pure $\mathcal{N}=4$, $D=4$ supergravity if all the non-$\mathrm{SU}(4)_{\textrm{S}}$-singlets inside $\tilde{X}$ belong to $K$. A direct evaluation shows that $\widetilde{X}$ is contained in the following irreps: $(\mathcal{A}_{1})_{\hat{\imath} \hat{\jmath}} \in (\mathbf{10},\mathbf{1})_{2} \subset  K$, $(\mathcal{A}_{1})_{ij} \in (\mathbf{1},\mathbf{10})_{-2} \subset W^{\textrm{int}}$, $(\mathcal{A}_{2})_{\hat{\imath}}{}^{\hat{\jmath}\hat{k} l} \in   (\mathbf{\bar{20}},\mathbf{\bar{4}})_{0}  \subset K$ and $(\mathcal{A}_{2})_{\hat{\imath}}{}^{\hat{\jmath} kl} \in  (\mathbf{15},\mathbf{6})_{2} \subset K $. Therefore, all the components of the intrinsic torsion $T^{\textrm{int}}_{\mathcal{N}=4}$ that are \textit{not} $\textrm{SU}(4)_{\textrm{S}}$-singlets are contained in $K$. Following \cite{Gallerati:2014xra}, one can show that this is a consequence of the constraints on the fermion shift tensors imposed by supersymmetry. This ensures the consistency of the unorthodox $\mathcal{N}=4$ subsector describing pure $\mathcal{N}=4$ supergravity with the scalars $(\chi,\xi)$ parameterising the coset space
\begin{equation}
\frac{\text{Comm}_{\text{E}_{7(7)}}(\textrm{SU}(4)_{\textrm{S}})}{\text{Comm}_{\text{SU}(8)}(\textrm{SU}(4)_{\textrm{S}})} = \frac{\textrm{SL}(2)}{\textrm{SO}(2)} \ .
\end{equation}  
The fermion shifts in the pure $\mathcal{N}=4$ $\textrm{SO}(4)_{\textrm{R}}$-gauged supergravity are identified with $(\mathcal{A}_{1})_{ij}$ and $(\mathcal{A}_{2})_{i}{}^{jkl}$ in (\ref{A1_tensor}) and (\ref{A2_tensor}), respectively \footnote{Up to normalisation factors that can be read off from the supersymmetry variations of the fermions: $(\mathcal{A}_1)_{ij} = \tfrac{\sqrt{2}}{3}(\mathcal{A}_1^{\mathcal{N}=4})_{ij}$ and $\tfrac{1}{6}(\mathcal{A}_2)_i{}^{klm}\epsilon_{klmj} = \tfrac{\sqrt{2}}{3}\, (\mathcal{A}_2^{\mathcal{N}=4})_{ij}$.}.

\section{Type IIB uplift of the $\textrm{U}(1)_{\textrm{R}}^2$-invariant sector}

The type IIB oxidation of the $\left[\textrm{SO}(6)_{\textrm{diag}} \times \textrm{SO}(1,1) \right] \ltimes \mathbb{R}^{12}$-gauged maximal supergravity can be performed using a generalised Scherk--Schwarz (gSS) ansatz based on the specific twist matrix $U$ constructed in \cite{Inverso:2016eet}. In order to uplift the unorthodox subsector describing pure $\mathcal{N}=4$ supergravity, we have to minimally modify the gSS ansatz of \cite{Inverso:2016eet}. In particular, the twist matrix $U$ that generates the embedding tensor $X$ of \cite{Inverso:2016eet} must be replaced by a new twist matrix
\begin{equation}
\widetilde{U} = U \cdot \mathcal{V}_{\textrm{vac}}(\Phi_{0}) \ ,
\end{equation}
that generates $\widetilde{X}$ in (\ref{X_new}). The twist matrix $\widetilde{U}$ contains the information about the gauged maximal supergravity of \cite{Inverso:2016eet} once it has been Higgsed and expanded about its $\mathcal{N}=4$ vacuum. Equivalently, one could uplift the $\left(\mathcal{V}_{\textrm{vac}}(\Phi_{0}) \cdot \textrm{SU}(4)_{\textrm{S}} \cdot \mathcal{V}_{\textrm{vac}}(\Phi_{0}){}^{-1}\right)$-invariant sector of the gauged maximal supergravity of \cite{Inverso:2016eet} using the original twist matrix $U$ therein.

As an illustration of the procedure, we will present the type IIB uplift on $\textrm{S}^{1} \times \textrm{S}^{5}$ of the simpler abelian $\textrm{U}(1)_{\textrm{R}}^2$-invariant sector of pure $\mathcal{N}=4$, $D=4$ supergravity. This supergravity model was investigated in \cite{Ferrero:2021ovq}, where a family of multi-dyonically charged and rotating supersymmetric $\textrm{AdS}_{2} \times \Sigma$ solutions, with $\Sigma$ being a spindle, was constructed. The uplift of such solutions to 11D supergravity on $\textrm{S}^{7}$ was also presented in \cite{Ferrero:2021ovq}. Upon a straightforward use of the formulae to be presented in this section, the $\textrm{AdS}_{2} \times \Sigma$ solutions of \cite{Ferrero:2021ovq} can alternatively be (locally) uplifted to S-fold backgrounds of type IIB supergravity on $\textrm{S}^{1} \times \textrm{S}^{5}$. Lastly, the type IIB embedding of the complete (non-abelian) pure $\mathcal{N}=4$, $D=4$ $\textrm{SO}(4)_\textrm{R}$-gauged supergravity will be presented in a follow-up paper.

\subsection{$\textrm{U}(1)_{\textrm{R}}^2$-invariant sector}

The $\textrm{U}(1)_{\textrm{R}}^2$-invariant sector of the $\mathcal{N}=4$, $D=4$ supergravity retains the two electric vectors gauging the Cartan subalgebra of $\textrm{G}_{\mathcal{N}=4}$, \textit{i.e.} $\mathfrak{u}(1)_{\textrm{R}}^{2} \subset \mathfrak{so}(4)_{\textrm{R}}$. Choosing, without loss of generality, such electric vectors to be
\begin{equation}
\label{A1&A2_def}
A_{1} \equiv A_{(1)}{}^{1} 
\hspace{6mm} \textrm{ and } \hspace{6mm}
A_{2} \equiv A_{(1)}{}^{\hat{1}} \ ,
\end{equation}
the bosonic action in (\ref{action_N=4}) precisely reduces to the supergravity model of \cite{Ferrero:2021ovq}. Despite not entering the action, it will be convenient to still consider the magnetic vectors $\tilde{A}_{1}$ and $\tilde{A}_{2}$ when it comes to present the type IIB uplift formulae. Electric and magnetic vectors do not carry independent dynamics as they obey a twisted self duality relation of the form
\begin{equation}
\label{twisted_SD_cond}
\star \mathcal{F} = - \mathbb{C} \, \mathcal{M} \, \mathcal{F} \ ,
\end{equation}
where $\mathcal{F} \equiv (F_{1},F_{2},\tilde{F}_{1},\tilde{F}_{2})$ with $F_{i}=dA_{i}$ and $\tilde{F}_{i}=d\tilde{A}_{i}$ ($i=1,2$), $\mathbb{C}$ is the skew-symmetric $\textrm{Sp}(4)$-invariant matrix, and
\begin{equation}
\label{calM_scalar}
\mathcal{M} = - e^\xi \left(\begin{array}{cccc}
1 & 0 & \chi & 0 \\
0 & |\tau|^2 & 0 & -\chi \\
\chi & 0 & |\tau|^2 & 0\\
0 & -\chi & 0 & 1
\end{array}\right) \ .
\end{equation}
From (\ref{twisted_SD_cond})-(\ref{calM_scalar}) one finds
\begin{equation}
\label{Ftilde_def}
\begin{array}{lcll}
\tilde{F}_{1} &=& |\tau|^{-2} \left( e^{-\xi} \star F_{1} - \chi \,  F_{1} \right)  \ ,  \\[4mm]
\tilde{F}_{2} &=&  e^{-\xi}  \star F_{2} + \chi \, F_{2} \ ,
\end{array}
\end{equation}
and the set of Bianchi identities and equations of motion for the electric vectors entering the action can be expressed in the compact form
\begin{equation}
\label{vector_eqs}
d\mathcal{F} = 0 \ .
\end{equation}
The equations of motion for the scalars are given by
\begin{equation}
\label{eom_scalars}
\begin{array}{rcl}
\square \xi - e^{2\xi} \partial_\mu \chi \partial^\mu \chi + \frac{1}{4} \mathcal{F}^{T}_{\mu\nu} (\partial_\xi \mathcal{M}) \mathcal{F}^{\mu\nu} - \partial_\xi V &=& 0 \ , \\[2mm]
\nabla_\mu (e^{2\xi} \partial^\mu\chi) + \frac{1}{4} \mathcal{F}^{T}_{\mu\nu} (\partial_\chi \mathcal{M}) \mathcal{F}^{\mu\nu} - \partial_\chi V &=& 0 \ ,
\end{array}
\end{equation}
whereas the Einstein field equations read
\begin{equation}
\label{Einstein_eq}
2\,R_{\mu\nu}=\partial_\mu \xi \partial_\nu \xi + e^{2\xi} \partial_\mu \chi\,\partial_\nu\chi-  \mathcal{F}^{T}_{\mu\rho} \mathcal{M} {\mathcal{F}_{\nu}}^{\rho} + V g_{\mu\nu} \ .
\end{equation}
The Bianchi identities and field equations (\ref{vector_eqs})-(\ref{Einstein_eq}) will be recovered from the type IIB ten-dimensional Bianchi identities and field equations upon use of the truncation ansatz in the next section. Finally, setting $\chi=\xi=0$ and identifying the vector fields as $A_1=A_2 \equiv A$ and $\tilde{A}_1=\tilde{A}_2 \equiv \tilde{A}$, one is left with the action and equations of motion of pure $\mathcal{N}=2$ gauged supergravity. Therefore, the type IIB formulae we are presenting can straightforwardly be used to uplift any solution of pure $\mathcal{N}=2$ gauged supergravity (see \cite{Caldarelli:2003pb} and references therein).

\subsection{Uplift to type IIB on $\textrm{S}^{1} \times \textrm{S}^{5}$}

We adopt the conventions of \cite{Guarino:2022tlw} in order to describe the $\textrm{S}^1 \times \textrm{S}^5$ internal geometry in the type IIB uplift of the $\textrm{U}(1)_{\textrm{R}}^2$-invariant sector. The $\textrm{S}^{1}$ is parameterised by a periodic coordinate $\eta \in [0,\,T]$ of period $T$ whereas the $\textrm{S}^{5}$ is understood as two $2$-spheres $\textrm{S}_{i}^{2}$, with $i=1,\,2$, with polar and azimuthal angles $(\theta_i,\,\varphi_i)$ fibered over an interval $\alpha \in [0,\,\frac{\pi}{2}]$. The $\textrm{S}^{5}$ is embedded in $\mathbb{R}^6$ using embedding coordinates
\begin{equation}
\label{embedding_coordinates}
\begin{array}{lcll}
    Y_1 = \cos\alpha\,\cos\theta_1 & , & Y_4 = \sin\alpha\,\cos\theta_2 & , \\
    Y_2 = \cos\alpha\,\sin\theta_1\,\sin\varphi_1 & , & Y_5 = \sin\alpha\,\sin\theta_2\,\sin\varphi_2 & ,  \\
    Y_3 = \cos\alpha\,\sin\theta_1\,\cos\varphi_1  & , & Y_6 = \sin\alpha\,\sin\theta_2\,\cos\varphi_2 & ,
\end{array}
\end{equation}
satisfying $\sum Y_{m}^{2}=1$. In order to lighten the type IIB uplift formulae, it will prove convenient to introduce the $(\tau,\alpha)$-dependent functions
\begin{equation}
\begin{array}{rcl}
f_0 &=& 1 + 2\, e^{-\xi}\, \cos^2\alpha \ , \\[2mm]
f_1 &=& 1 + 2\, e^\xi \,|\tau|^{2} \, \cos^2\alpha \ , \\[2mm]
f_2 &=& 1 + 2\, e^\xi \,\sin^2\alpha \ . 
\end{array}
\end{equation}

Equipped with the various functions and definitions above, the ansatz for the ten-dimensional metric takes the form
\begin{equation}
\label{10D_metric}
ds_{10}^2 = \Delta^{-1} \left(\tfrac{1}{2}\, ds_{4}^2 + g_{mn} Dy^m Dy^n \right) \ ,
\end{equation}
where $y^m=(\eta,\,\alpha,\,\theta_i,\,\varphi_i)$ denotes the internal coordinates and $D \equiv d+A^{\textrm{KK}}$ becomes covariantised with respect to the Kaluza--Klein (KK) vector
\begin{equation}
\label{KK_vector}
A^{\textrm{KK}} \equiv A_{\mu}{}^{m} dx^{\mu } \otimes \partial_{m} = A_1\,\partial_{\varphi_1}+ A_2 \,\partial_{\varphi_2} \ .
\end{equation}
The four-dimensional vectors $A_{1}$ and $A_{2}$ entering (\ref{KK_vector}) precisely correspond to the ones in (\ref{A1&A2_def}). The metric $g_{mn}$ on the internal $\textrm{S}^{1} \times \textrm{S}^{5}$ is given by
\begin{equation}
\label{g_internal}
g_{mn} \, dy^{m} dy^{n} = d\eta^2  + d\alpha^2  + \dfrac{\cos^2\alpha}{f_1} ds_{\textrm{S}^2_1}^2+ \dfrac{\sin^2\alpha}{f_2} ds_{\textrm{S}^2_2}^2 \ ,
\end{equation}
with
\begin{equation}
ds_{\textrm{S}^2_i}^2  = d\theta_{i}^2 + \sin^2\theta_{i} \, {d\varphi_{i}}^2
\end{equation}
being the line element of the $2$-spheres $\textrm{S}^2_i$ of unit radius. There is also a non-singular warping factor 
\begin{equation}
\Delta^{-4} = f_1 \, f_2 \ .
\end{equation}

The ansatz for the type IIB dilaton $\Phi$ and the RR axion $C_{0}$ is given by
\begin{equation}
\label{Phi&C0}
e^\Phi = \Delta^2 \, e^{-2 \eta} \, e^{\xi}\, f_0
\hspace{2mm} , \hspace{3mm} 
C_{0} = e^{2\eta} \,\chi \, \cos(2\alpha)\,f_0{}^{-1}.
\end{equation}
The two-form potentials $\mathbb{B}^{\alpha}=(\mathbb{B}^{1},\mathbb{B}^{2}) = (B_{2},C_{2})$ have a decomposition \`a la KK of the form
\begin{equation}
\label{B_definitions}
\mathbb{B}^\alpha = e^{(-1)^\alpha \eta} \,\mathfrak{b}^\alpha_{(0)} + \mathfrak{b}^\alpha_{(1)} \ ,
\end{equation}
involving four-dimensional scalar contributions
\begin{equation}
\label{B2&C2_scalar_contrib}
\begin{array}{rcl}
\mathfrak{b}^1_{(0)} &=& - \sqrt{2}\,  \chi\,e^{\xi}\, \cos\alpha\,\,\widetilde{\textrm{vol}}_{1} - \sqrt{2}\, (1+e^{\xi})\,\sin\alpha\, \,\widetilde{\textrm{vol}}_{2} \ , \\[2mm]
\mathfrak{b}^2_{(0)} &=& \sqrt{2}\,(1+ e^\xi |\tau|^2) \, \cos\alpha\, \,\widetilde{\textrm{vol}}_{1} - \sqrt{2}\, \chi\, e^\xi\,\sin\alpha\,\widetilde{\textrm{vol}}_{2} \ ,
\end{array}
\end{equation}
given in terms of \emph{fibered} $2$-sphere volumes
\begin{equation}
\label{twisted_volumes}
\begin{array}{rcll}
\widetilde{\textrm{vol}}_{1} &=& \cos^2\alpha\,f_1{}^{-1} \, \sin{\theta_1}\, d\theta_1 \wedge D\varphi_1 & , \\[2mm]
\widetilde{\textrm{vol}}_{2} &=& \sin^2\alpha\,f_2{}^{-1} \,\sin{\theta_2} \, d\theta_2 \wedge D\varphi_2 & ,
\end{array}
\end{equation}
and one-form contributions
\begin{equation}
\label{B2&C2_vector_contrib}
\begin{split}
   & \mathfrak{b}^1_{(1)} = \tfrac{1}{\sqrt{2}} \, \tilde{A}_1 \, D(-e^{-\eta}\, Y_1) + \tfrac{1}{\sqrt{2}} \, A_2 \, D(e^{-\eta}\, Y_4) \ , \\[2mm]
    &\mathfrak{b}^2_{(1)} = \tfrac{1}{\sqrt{2}} \, A_1 \, D(-e^{\eta}\, Y_1) + \tfrac{1}{\sqrt{2}} \, \tilde{A}_2 \, D(e^{\eta}\, Y_4) \ .
\end{split}
\end{equation}
The one-form contributions (\ref{B2&C2_vector_contrib}) are concisely expressed in terms of the four-dimensional electric $A_{i}$ and magnetic $\tilde{A}_{i}$ vectors whose field strengths $F_{i}$ and $\tilde{F}_{i}$ are related by the twisted self-duality condition (\ref{twisted_SD_cond}) as given in (\ref{Ftilde_def}). Both $F_{i}$ and $\tilde{F}_{i}$ will appear when computing the three-form field strengths $\mathbb{H}^{\alpha}=d\mathbb{B}^{\alpha}$.

In order to present the type IIB self-dual five-form $\widetilde{F}_{5}=dC_{4} + \frac{1}{2} \, \epsilon_{\alpha \beta} \, \mathbb{B}^{\alpha}\wedge \mathbb{H}^{\beta}$, let us introduce the quantities
\begin{equation}
\begin{array}{llll}
v_1 &=& \chi \, e^\xi \, Y^1\, e^{-\eta} \, d(e^{\eta}\cos\alpha )\wedge \widetilde{\textrm{vol}}_{1} & , \\[2mm]
v_2 &=& \chi \, e^\xi \, Y^4\, e^{\eta} \, d(e^{-\eta}\sin\alpha)\wedge \widetilde{\textrm{vol}}_{2} & ,
\end{array}
\end{equation}
and
\begin{equation}
\begin{array}{lll}
\tilde{v}_1&=& (1+ e^\xi |\tau|^2) Y^1 e^\eta d(e^{-\eta} \cos\alpha) \wedge \widetilde{\textrm{vol}}_{1} \\[2mm]
&& + \, \frac{1}{2} \, d((Y^2)^2  + (Y^3)^2) \wedge D\varphi_1\wedge d\eta \ , \\[4mm]
\tilde{v}_2 &=& -(1+ e^\xi) Y^4 e^{-\eta} d(e^{\eta} \sin\alpha) \wedge \widetilde{\textrm{vol}}_{2} \\[2mm]
&& + \, \frac{1}{2} \, d((Y^5)^2  + (Y^6)^2) \wedge D\varphi_2\wedge d\eta \ .
\end{array}
\end{equation}
In terms of them, the field-strength $\widetilde{F}_{5}$ can then be expressed as
\begin{equation}
\label{tildeF_5}
\begin{array}{lll}
\widetilde{F}_5 &=&-(1+\star)\Big[\Big( (2+V) \sin(2\alpha) \, d\eta \\
&& + 2 \big( (2+e^\xi) \cos(2\alpha) + V \cos^2 \alpha \big) d\alpha \Big) \wedge \widetilde{\textrm{vol}}_{1} \wedge \widetilde{\textrm{vol}}_{2} \\[2mm]
&& + \sin(2\alpha) \left( e^{2\xi}\, \chi \,d\chi - d\xi\right)\wedge \widetilde{\textrm{vol}}_{1} \wedge \widetilde{\textrm{vol}}_{2}  \\[2mm]
&& + F_1 \wedge v_1 + \tilde{F}_1 \wedge \tilde{v}_1 + F_2 \wedge v_2 + \tilde{F}_2 \wedge \tilde{v}_2 \Big] \ ,
\end{array}
\end{equation}
where the ten-dimensional self-duality of (\ref{tildeF_5}) requires the four-dimensional twisted self-duality condition (\ref{twisted_SD_cond}), equivalently, (\ref{Ftilde_def}).

We have verified that, when plugging the above reduction ansatz into the ten-dimensional equations of motion and Bianchi identities of type IIB supergravity, then the four-dimensional equations (\ref{vector_eqs})-(\ref{Einstein_eq}) are recovered provided the self-duality condition (\ref{twisted_SD_cond}) holds \footnote{The Einstein equations have been verified using some of the non-trivial examples in \cite{Ferrero:2021ovq}}. This guarantees the consistency of the truncation of type IIB supergravity on $\textrm{S}^{1} \times \textrm{S}^{5}$ down to the $\textrm{U}(1)_{\textrm{R}}^2$-invariant sector of $\mathcal{N}=4$, $D=4$ supergravity.  Finally, the type IIB uplift we have presented admits a pair of axion-like deformations $\chi_{i=1,2}$ along the lines of \cite{Guarino:2021hrc}. These can break supersymmetry and enter the uplift formulae via the simple replacement $d\varphi_{i} \rightarrow d\varphi_{i} +  \chi_i \,d\eta$.

\subsection{S-fold interpretation}

Let us look in detail at the dependence of the various type IIB fields on the coordinate $\eta$ along the $\textrm{S}^{1}$. This dependence is intimately related to the transformation properties of the type IIB fields under the (classical) S-duality group $\textrm{SL}(2,\mathbb{R})$. The ten-dimensional metric (\ref{10D_metric}) and the field strength $\widetilde{F}_{5}$ (\ref{tildeF_5}) are S-duality singlets and, consistently, do not depend on $\eta$. On the contrary, the two-form potential $\mathbb{B}^{\alpha}=(B_{2},C_{2})$ and the type IIB axion-dilaton matrix
\begin{equation}
m_{\alpha\beta} = \begin{pmatrix} e^{-\Phi} + e^{\Phi} \, C_{0}^2 & - e^{\Phi} \, C_{0} \\  - e^{\Phi} \, C_{0} & e^{\Phi}  \end{pmatrix} \ ,
\end{equation}
transform linearly under S-duality. By looking at (\ref{B_definitions})-(\ref{B2&C2_vector_contrib}) and (\ref{Phi&C0}), one observes a very specific dependence of these fields on the coordinate $\eta$. In particular, it is totally encoded in a (local) $\textrm{SL}(2,\mathbb{R})$ twist matrix
\begin{equation}
\label{A_matrix}
A^{\alpha}{}_{\beta}(\eta) = 
\begin{pmatrix}
e^{-\eta} &  0  \\
0  &  e^{\eta}
\end{pmatrix} \ ,
\end{equation}
of hyperbolic type, in terms of which the two-form potentials and the axion-dilaton matrix can be written as
\begin{equation}
\label{B&m_twist}
\mathbb{B}^{\alpha}=A^{\alpha}{}_{\beta}\,\mathfrak{b}^{\beta}
\hspace{2mm} \textrm{ and } \hspace{2mm}
m_{\alpha\beta} = (A^{-t})_{\alpha}{}^{\gamma} \, \mathfrak{m}_{\gamma\delta}  (A^{-1})^{\delta}{}_{\beta}
\end{equation}
with $\mathfrak{b}^{\beta}$ and $\mathfrak{m}_{\gamma\delta}$ being $\eta$-independent fields. 

The coordinate $\eta$ can be made periodic of period $T$. However, in this case, the $\textrm{SL}(2,\mathbb{R})$ twist matrix in (\ref{A_matrix}) induces a non-trivial monodromy
\begin{equation}
\label{monodromy}
\mathfrak{M}_{\textrm{S}^{1}} = A^{-1}(\eta) \, A(\eta + T) =  \begin{pmatrix} e^{-T} & 0\\ 0 & e^{T} \end{pmatrix} \ ,
\end{equation}
on the type IIB fields (\ref{B&m_twist}) upon which it acts. In order for a background of this kind to be a consistent S-fold solution of type IIB superstring theory, the monodromy along the S$^1$ should belong to the (quantum) S-duality group $\textrm{SL}(2,\,\mathbb{Z})$. This is achieved by acting on the above solution with a suitable (global) element $g\in \textrm{SL}(2,\mathbb{R})$, as a consequence of which, the monodromy matrix becomes
\begin{equation}
\mathfrak{M}(k) = g^{-1} \, \mathfrak{M}_{\textrm{S}^{1}} \, g =
\begin{pmatrix}
    k & 1\\
    -1 & 0 
\end{pmatrix}  
\hspace{2mm} \textrm{ with }\hspace{2mm}
g \in \textrm{SL}(2,\mathbb{R})\ ,
\end{equation}
where $k \in \mathbb{N}$ and $k \geq 3$, provided $T$ is chosen to be $k$-dependent and given by $T(k) = \textrm{arccosh}\left(\frac{k}{2}\right)$. The non-trivial S-duality monodromy $\mathfrak{M}(k)$ renders the type IIB background locally geometric but globally non-geometric.

\section{Conclusions}

In this letter we have presented a new method to construct consistent truncations of type IIB and 11D supergravities based on Generalised Geometry and Exceptional Field Theory. Starting from an AdS$_{D}$ vacuum preserving $\mathcal{N}$ supersymmetries in a $D$-dimensional gauged maximal supergravity that admits a (locally) geometric uplift to type IIB/11D supergravity, the method provides a consistent truncation of the latter to pure $\mathcal{N}$-extended supergravity in $D$ dimensions. As an example, starting from the unique $\mathcal{N}=4$ AdS$_{4}$ vacuum of the $D=4$ gauged maximal supergravity of \cite{Inverso:2016eet}, which admits a non-geometric (still locally geometric) S-fold uplift to type IIB supergravity, we have constructed a consistent truncation of the latter down to pure $\mathcal{N}=4$, $ D=4$ $\textrm{SO}(4)_{\textrm{R}}$-gauged supergravity. We can then state that:
\\[-3mm]

\noindent\textit{``Any solution of pure $\mathcal{N}=4$, $D=4$ $\textrm{SO}(4)_{\textrm{R}}$-gauged supergravity can be oxidised to a non-geometric S-fold background of type IIB string theory."}
\\[-3mm]

\noindent Particularly interesting examples of such four-dimensional solutions are the multi-charge accelerating black holes and spindles of \cite{Ferrero:2021ovq} which have already been embedded in M-theory using the uplift formulae provided in \cite{Cvetic:1999au}. Having both a geometric M-theory embedding and a non-geometric type IIB embedding at disposal for such supergravity solutions opens up the possibility for a comparative holographic study within the contexts of ABJM theory \cite{Aharony:2008ug} (\textit{cf.} \cite{Colombo:2024mts}) and S-fold CFT$_{3}$'s \cite{Assel:2018vtq} (see also \cite{Lozano:2016wrs} for a connection between M-theory and type IIB supergravity solutions based on non-abelian T-duality). This could help in shedding new light on the holography of non-geometric backgrounds of string theory.

Although we have focalised our discussion on a \textit{pure} $\mathcal{N}=4$ supergravity subsector of the maximal gauged supergravity of \cite{Inverso:2016eet}, our method holds on more general grounds. Firstly, it does not require the consistent subsector of the maximal gauged supergravity to be a pure $\mathcal{N}$-extended supergravity. A follow-up application in this direction would be to assess the existence of a consistent subsector of the maximal gauged supergravity of \cite{Inverso:2016eet} describing $\mathcal{N}=2$ supergravity coupled to a massless vector multiplet and capturing the AdS$_{4}$ vacuum dual to a generic point in the conformal manifold of $\mathcal{N}=2$ S-fold CFT$_{3}$'s \cite{Bobev:2021yya} (which comes along with a $\textrm{U}(1)_{\textrm{F}}$ flavour symmetry). Our method would automatically provide the relevant type IIB uplift formulae for such a matter-coupled $\mathcal{N}=2$ consistent truncation (simpler than the one of \cite{Guarino:2024zgq}), which could then be used to perform precision tests of S-fold holography building on the results of \cite{Bobev:2023bxs}. Secondly, and upon suitable adjustments, our method can also be applied to identify consistent subsectors of less supersymmetric and matter-coupled gauged supergravities for which a (locally) geometric higher-dimensional origin, as well as an embedding tensor formulation, still exist.

As originally conjectured in \cite{Gauntlett:2007ma}, a consistent truncation of type IIB supergravity down to pure $\mathcal{N}=4$, $D=4$ supergravity must exist for each of the $\textrm{AdS}_{4} \ltimes \textrm{K}$ solutions in \cite{Assel:2011xz,Assel:2012cj}. These are in turn obtained as special limits of the Janus solutions in \cite{DHoker:2007zhm,DHoker:2007hhe} (see \cite{Akhond:2021ffz} for an alternative description of the solutions in terms of a Laplace problem), and are conjectured to holographically describe infrared fixed points of a large class of $\mathcal{N}=4$ CFT$_{3}$'s of Gaiotto--Witten type \cite{Gaiotto:2008ak}. One of the simplest solutions corresponds to the $\mathcal{N}=4$ S-fold of \cite{Inverso:2016eet}, which furthermore is the \textit{unique} $\mathcal{N}=4$ AdS$_{4}$ vacuum of a $D=4$ gauged maximal supergravity with a known higher-dimensional origin \cite{Gallerati:2014xra,Inverso:2016eet}. Inspired by the type  IIB reduction ansatz provided in this letter, it would be interesting to investigate possible modifications thereof yielding the pure $\mathcal{N}=4$, $D=4$ consistent truncation capturing a generic solution of the type in \cite{DHoker:2007zhm,DHoker:2007hhe,Assel:2011xz,Assel:2012cj}. This result would be instrumental in the holographic study of more general classes of type IIB solutions approaching \cite{DHoker:2007zhm,DHoker:2007hhe,Assel:2011xz,Assel:2012cj} asymptotically.

Finally, it would be interesting to explore whether the consistency of a subsector of a $D$-dimensional gauged supergravity can be established without reference to a higher-dimensional theory. As we have shown, the proof of consistency only relies on a group-theoretic analysis of the embedding tensor describing the $D$-dimensional gauged supergravity. However, a key conceptual step in the proof is the identification of the embedding tensor with an intrinsic torsion, which is, of course, meaningless if no higher-dimensional origin exists (as proved in \cite{Lee:2015xga,Inverso:2017lrz} for the $\mathrm{SO}(8)_{\omega\neq 0}$ gaugings of \cite{DallAgata:2012mfj}). We leave these and other related questions for the future.

\vspace{5mm}

\noindent\textbf{Acknowledgements}:  
We are grateful to Carlos N\'u\~nez, Gianluca Inverso, Emanuel Malek, Anik Rudra and Daniel Waldram for interesting conversations. The work of AG is supported by the Spanish national grant MCIU-22-PID2021-123021NB-I00. The work of C.S. has been supported by an INFN postdoctoral fellowship, Bando 24736.

\bibliographystyle{utphys}
\bibliography{references}

\end{document}